\documentclass[aps,prl,letterpaper,11pt,twoside,tightenlines,nofootinbib,showpacs,preprint,twocolumn]{revtex4}
\usepackage{graphicx}
\usepackage[sort&compress]{natbib}
\usepackage{latexsym}
\usepackage{epsfig}
\begin{document}
\newcommand{\eg}{{\it e.g.}}
\newcommand{\etal}{{\it et. al.}}
\newcommand{\ie}{{\it i.e.}}
\newcommand{\be}{\begin{equation}}
\newcommand{\dd}{\displaystyle}
\newcommand{\ee}{\end{equation}}
\newcommand{\bea}{\begin{eqnarray}}
\newcommand{\eea}{\end{eqnarray}}
\newcommand{\bef}{\begin{figure}}
\newcommand{\eef}{\end{figure}}
\newcommand{\bce}{\begin{center}}
\newcommand{\ece}{\end{center}}
\def\lsim{\mathrel{\rlap{\lower4pt\hbox{\hskip1pt$\sim$}}
    \raise1pt\hbox{$<$}}}         
\def\gsim{\mathrel{\rlap{\lower4pt\hbox{\hskip1pt$\sim$}}
    \raise1pt\hbox{$>$}}}         

\title{Holographic Bound in Quantum Field Energy Density and Cosmological Constant}
\author{P. Castorina}
\affiliation{Dipartimento di Fisica, Universit\`a di Catania,
 and INFN Sezione di Catania, Via Santa Sofia 64
I-95123 Catania, Italia}

\date{\today}
\begin{abstract}
The cosmological constant problem is reanalyzed by imposing the limitation of the number of degrees of freedom (d.o.f.) due to entropy bounds directly in the calculation of the energy density of a field theory. It is shown that if a quantum field theory has to be consistent with gravity and holography, i.e. with an upper limit of storing information in a given area, the ultraviolet momentum cut-off is not the Planck mass, $M_p$, as naively expected, but $M_p/N_U^{1/4}$ where $N_U$ is the number of d.o.f. of the universe. The energy density evaluation turns out completely consistent with Bousso's bound on the cosmological constant value. The scale $M_p/N_U^{1/4}$, that in the "fat graviton" theory corresponds to the graviton size, originates by a self-similar rearrangement of the elementary d.o.f. at different scales that can be seen as an infrared-ultraviolet connection.
 
\end{abstract}
 \maketitle

\section{Introduction}

In 1955 von Weizsaker \cite{weiz} started a program to derive quantum theory by postulating a fundamental quantized binary alternative, called "ur",
and, many years later , Wheeler \cite{whel} suggested that information theory must play a relevant role in understanding the foundations of quantum mechanics, the "It for bit" proposal.

On the other hand, it is now generally accepted that the picture of space-time as a locally flat Minkowsky manifold breaks down at distances of order of Planck scale, $l_p \simeq 1.6 *10^{-33}$ cm, and that, due to space-time uncertainty  \cite{pen,dop}, an elementary area $A_p = l_p^2$  should be considered as a fundamental space-time cell. 

The previous  ideas are the starting points of the main approaches to quantum gravity that consider quantum space-time to be discrete:
loop quantum gravity  \cite{rovelli} and string theory \cite{string}.

Moreover , by combining the assumptions of a quantized fundamental (binary) degree of freedom  (d.o.f.) and of an elementary area, $A_p$, with Bekenstein's result on black-hole entropy \cite{beck1}, the Holographic Principle emerges \cite{hooft,suss,suss2,wit,mal,beck2,fw1,bousso1} :

a) the number of possible states in a region of space is the same as that of a system of binary d.o.f. distributed on the causal horizon  of the region;

b) the number of d.o.f., $N$, of a region of space is bounded by the area , $A$, in unit of the elementary area $A_p$ ($c= \hbar =1$):
\be
 N \le \frac{A}{A_p} 
\ee

Holography has a crucial role in the description of gravity as an emergent phenomenon of thermodynamical origin \cite{jacobson, thanu,pad2}
and/or as an entropic force \cite{verlinde}.

Moreover , following the proposal in \cite{bank,fs2} that the cosmological constant, $\Lambda$,   should be related to the number of d.o.f. in the fundamental theory, it has been shown \cite{bousso1} that in any universe with a positive cosmological constant one obtains, by holographic entropy bounds, \cite{beck2,bousso1}  an upper limit to the number of d.o.f. of the universe , $N_U$, given by  
\be
N_U \le \frac{3\pi}{l_p^4 \Lambda}  
\ee
i.e.
\be
\Lambda \le  \frac{3\pi}{l_p^4 N_U}
\ee

The problem  \cite{wein} is to understand how the result in eq.(3) can be (re-)obtained from the point of view of a field theory where
the naive estimate of the energy density, with $M_p =1/l_p$ as ultraviolet cut-off, is given by
\be
\Lambda  = \int^{M_p} \frac{d^3 k}{(2\pi)^3} k \simeq M_p^4/8\pi^2 = 1/8 \pi^2 l_p^4.
\ee

A direct comparison between this equation and  eq. (3) clearly indicates that to reproduce the bound in eq. (3) one has to introduce  a limitation of the number of d.o.f. in the quantum field estimate of the energy density.  Indeed, standard local field theories overcount available degrees of freedom because they fail to include the effect of gravitation \cite{bousso1} and the discrepancy is due to number of d.o.f. of the universe which, according to the holographic principle, has an upper bound given by  the number of elementary cells on a spherical causal horizon of area, $A_U$, with radius equal to the inverse Hubble constant $H$ ($H^{-1} \simeq 2.7 * 10^{61} l_p$), that is
\be
N_U \le \frac{A_U}{l_p^2} \simeq \frac{1}{H^2 l_p^2}\simeq 10^{122}.
\ee

In the absence of a unified theory of gravity and quantum fields, in this letter  a ,"crude", new method is proposed to include  in the calculation of the energy density of a field theory the limitation of the number of d.o.f. in agreement with the entropy bound.

\section{Holographic bound in quantum field energy density}

An interesting attempt in understanding the value of the cosmological constant in particle physics has been done \cite{coen} by considering the energy - not the energy density - of a quantized field in a box of size $L$. The energy is (volume*energy density) and therefore it is of order $L^3 M_p^4$. By assuming that the lagrangian of the theory describes all state of the system excluding those for which it has already collapsed to a black-hole, a relation between the infrared cut-off, $L$, and the ultraviolet cut-off, $M_p$, arises in such a way that the cosmological constant turns out
\be
\Lambda \simeq (l_pL)^{-2}
\ee
which gives ( within an order of magnitude) the observed value of $\Lambda$ if one identifies $L$ with the Hubble constant ( see ref. \cite{coen} for details).

However \cite{hsu}, since $\Lambda \simeq 1/L^2$ and in the standard cosmological model, in a dominated matter era, $L \simeq R(t)^{3/2}$ ($R(t) $ is the scale factor) , then $\Lambda \simeq 1/R(t)^3$ which corresponds to an equation of state with $w=0$ ($w=p/\rho$ where $p$ is the pressure and $\rho$ the energy density of the system) rather than the value $w \simeq -1$  required  for the dark energy by the analysis of cosmological data 
\cite{data1,data2,data3,data4}.

Let us rather consider the evaluation of the energy density of a quantum field by imposing the holographic entropy bound, valid for any physical system.

In general, if one covers an  area $A$ by elementary cells of area $A_p$, the number of the d.o.f. in the considered area is  $ \le A/A_p$. 

For a generic scale related to a momentum $k$, $l_k = 1/k$ and $l_k > l_p$, the previous limit on the number of d.o.f of the cell of area $l_k^2$ is
\be
N_k \le \frac{l_k^2}{A_p}
\ee
therefore
\be
k \le \frac{1}{l_p \sqrt{N_k}}
\ee

Now let us assume that 
\be
l_k^2 N_k \ge l_p^2 N_U
\ee
which is a non trivial point, that will be discussed in the next section, since it is a condition among different scales and different numbers of d.o.f. in
cells with very different sizes. By previous equation one has
\be
\sqrt{N_k} \ge  k l_p \sqrt{N_U}.
\ee
and  by eq.(8) 
\be
k \le \frac{1}{l_p N_U^{1/4}}
\ee
which implies that the  ultraviolet cut-off to take  correctly into account the holographic entropy bound is $1/(l_p N_U^{1/4})$ and not 
$1/l_p$ as naively expected.

Therefore the energy density of a "free" field theory turns out
\be
\Lambda \simeq \int ^{1/(l_p N_U^{1/4})} \frac{d^3k}{(2\pi)^3} k \simeq \frac{1}{8 \pi^2 l_p^4  N_U}
\ee
in agreement with the entropic bound obtained by Bousso \cite{bousso1} and with the experimental value of the cosmological constant 
($ \Lambda =(2.6 * 10^{-3})^4$ eV $^4$) \cite{data1,data2,data3,data4}. The previous ultraviolet cut-off is a direct consequence of gravitation. It disappears for $l_p \rightarrow 0$ and, as we shall discuss later, in a consistent calculation of the energy density for interacting fields, it should be interpreted as the typical momentum scale in the loop expansion of the gravitational effective action when standard model particles couple with external graviton legs. 

\section{Scaling behavior of fundamental degrees of freedom}

Let us now analyze the meaning of the  crucial assumption in eq.(9) and  let us initially  consider a geometrical argument by using, for semplicity, squares rather than spherical surfarces. The covering of a square of area $A=L^2$ with elementary squares is an old mathematical problem \cite{chung1}. The minimum number $\nu$ of elementary squares, of area $r^2$, needed to cover a large square of side length $L$ is given by \cite{chung2}
\be
\nu = \frac{L^2}{r^2} [ 1 + O((L/r)^{-(11-\sqrt{2})/7} log (L/r)]
\ee
where the positive second term in bracket is the "extra space". Therefore, in our case if one covers the entire horizon of the universe, $A_U$, with elementary cells of area $l_k^2$ one has
\be
\nu_k l_k^2 \ge A_U
\ee
where $\nu_k$ is the corresponding minimum number of cells to cover the horizon. 

Equation (9) is equivalent to assume that at any scale $l_k$ the number of d.o.f. $N_k$ ( in the cell of area $l_k^2$)  rearranges in such a way to be larger than or equal to to the minimum number of cell $\nu_k$  to cover the whole area, i.e. 
\begin{equation}
N_k \ge \nu_k.
\end{equation}
 Indeed, by previous eqs.(14,15) one gets
\be
N_k l_k^2 \ge \nu_k l_k^2 \ge A_U \ge l_p^2 N_U,
\ee    
which gives eq.(9).

Note that since $N_k$ and $N_U$ are integers, if $l_k$ is a multiple of $l_p$ and  one imposes the exact covering of the entire area ( no "extra space"), the condition $N_k=\nu_k$ corresponds to a self-similar rearrangement of d.o.f at different scales.
From this point of view, the consistency between the holographic entropy bound for the cosmological constant and the estimate of the energy density requires a peculiar self-similar rearrangement of the elementary d.o.f at different scales that can be seen as an infrared-ultraviolet connection. 

Let now discuss a different, more tradictional, approach to the meaning of the condition in eq.(11) which can be traslated, for example, in a gaussian type
cut-off  $exp(-k^2 l_p^2 N_U^{1/2})$ ( of course an exponential cut-off works equally well).

This implies that there is a typical scale in the field theory effectively coupled with gravity. In  the "fat graviton" theory \cite{sundrum1,sundrum2} 
this scale coincides with the effective size, $l_{gr}$, of the graviton and the local graviton coupling with the standard model (SM) particles is strongly modified. Infact, the dominant contributions to the gravitational effective action come from purely SM loops, with graviton external legs, which contribute only for wavelengths $> l_{gr}$. More precisely, with fat gravitons, massive SM (i.e. particle with mass $m_{SM} >> 1/l_{gr}$) and hard light SM pieces of loop contributions to the gravitational effective action may consistently \cite{sundrum1} be suppressed while the  soft light SM contribution are not, i.e.
there is no robust contribution to the cosmological constant from momentum scale larger than $1/l_{gr}$. However among the diagrams contributing to the cosmological constant there is one with no graviton external legs, corresponding to the free energy density, and the graviton size is unable to suppress this contribution. As discussed in detail in ref. \cite{sundrum1}, the cosmological term in the loop expansion of the gravitational effective
action is a self-interaction of the graviton field and, by invoking the general coordinate covariance , diagrams with soft graviton external lines are related to the diagram with no graviton. From this point of view, a contribution to the cosmological constant such as
\begin{equation}
\Lambda \simeq \int \frac{d^3 k}{(2\pi)^3} k \exp^{-k^2 l_{gr}^2}
\end{equation}
has to be interpreted as a short-hand for contributions from diagrams with gravitons interactions.

In this respect, the cut-off in eq.(11), although obtained by general consideration on entropy bounds, gives the typical momentum scale  in the loop expansion of the gravitational effective action when SM particles couple with external graviton legs.

The meaning of the graviton size comes by combining the two previous different points of view ,i.e. by imposing $l_{gr} = l_p N_U^{1/4}$. Indeed,
by entropy bound, the number of fundamental d.o.f. in a single cell of area $l_{gr}^2$, $N_{gr}$, 
turns out to be
\begin{equation} 
N_{gr} \le \frac{l_{gr}^2}{l_p^2}= N_U^{1/2}.
\end{equation}
On the other hand, the number of cell of size $l_{gr}$ one needs to cover the entire horizon is  $\nu_{gr} = A/l_{gr}^2$  and it turns out 
\begin{equation}
\nu_{gr} = \frac{A}{l_{gr}^2} \ge \frac{l_p^2 N_U}{l_p^2 N_U^{1/2}} = N_U^{1/2}
\end{equation}
By previous eqs. (18,19), the crucial condition in eq.(15), i.e.  $N_{gr} \ge \nu_{gr}$,  can be satisfied only by imposing $N_{gr}= \nu_{gr}$, that is the size of the graviton is such that the number of cells of area $l_{gr}^2$ one needs to cover the whole horizon is exactly equal to the number of fundamentl d.o.f.
$N_{gr}$ per cell.

\section{Comments and Conclusions}

It is interesting to note that eq.(5) can be obtained from an analysis of the representations of direct product of $SU(2)$ groups, each group related with an elementary binary system \cite{gor} and this clarifies the connection between the results in the previous sections and elementary quantum binary degrees of freedom.

There is another crucial aspect that has to be stressed. The final results in eqs.(11,12) do not require the saturation of the bounds in eqs.(2,8-11): the cosmological constant and the number of d.o.f. turn out to have upper limits. If one applies the saturated limits, the cosmological constant would be $\Lambda \simeq (1/L l_p )^{2}$ with the wrong equation of state, as  previously discussed, and one could obtain inconsistent results.

Rather $N_U$ should be considered as the maximum number of d.o.f. of a causal horizon if one uses elementary cells of size $l_p$ , but any other
elementary cell size can be used in discussing the entropy bound. Eq.(11) is  based on holographic entropy bound and on the crucial condition in eq.(9) and its meaning is that, because of gravity, not all d.o.f. that a field theory apparently supplies can be used for consistently storing information.

Finally, since the cosmological constant is directly related to the zero point energy density, one could comment by using statements related to Casimir effect. Indeed, in discussions of the cosmological constant, the Casimir effect is often invoked as decisive to rule out the possibility of an ultraviolet cut-off which gives result in accord with the experimental value. 

However, Casimir effects can be formulated and Casimir forces can be computed without reference to zero point energies \cite{bob}.
They are relativistic, quantum forces between charges and currents. The Casimir force (per unit
area) between parallel plates vanishes as $\alpha$, the fine structure constant, goes to zero, and the standard
result, which appears to be independent of $\alpha$, corresponds to the $\alpha \rightarrow \infty$ limit.

More generally, the physical role of the zero point energy is still an open problem  and  recent claims that vacuum fluctuactions of the electromagnetic field could be detected by experiments by Josephson junctions \cite{beck} have been definetely criticized in ref. \cite{enzo}.

In conclusion, if a quantum field theory has to be consistent with gravity, i.e. with an upper limit of storing information in a given area, there is an ultraviolet cut-off in the field mode of momentum $k$  given by eq.(11). The physical meaning of this cut-off in the fat graviton theory is the graviton size whereas in the approach proposed in this letter it originates by a self-similar behavior of the fundamental d.o.f.:
 the only "gravitating" modes are such that by covering the whole area of the system with the minimum number, $\nu_k$, of elementary cells of size $l_k=1/k$, the number of d.o.f. per cell, $N_k$, has to be larger than or equal to  $\nu_k$. In an effective field theory of SM particles interacting with gravitons, the cut-off $M_p/N_U^{1/4}$ has to be interpreted as the typical momentum scale in the loop expansion of the gravitational effective action when SM particles couple with external graviton legs.
It could be quite possible to have some physical effects at a scale $ \simeq 1/l_p N_U^{1/4}$ and  experiments on the possible modification of gravity
are very close to study this range of distances ( see for example \cite{new1,new2}).

The cosmological constant problem is far from to be solved \cite{carroll,nobb} and it is deeply related with quantum gravity, however the obtained results clearly indicate a new kind of infrared-ultraviolet connection that is worth to be investigated.

\section{Acknowledgements}
The author thanks M.Consoli and D.Zappala' for useful comments and suggestions and the Theoretical Physics Department of Bielefeld University for hospitality.


\begin{thebibliography}{99}
\bibitem{pad2} T.Padmanabhan , "Emergent perspective of Gravity and Dark energy" , arxive: 1207.0505, where the holographic equipartition of energy among
fundamental d.o.f. has a crucial role in understanding gravity as an emergent phenomenon.
\end{thebibliography}
\end{document}